\pdfoutput=1

\documentclass[conference]{IEEEtran}
\IEEEoverridecommandlockouts
\IEEEpubid{\begin{minipage}{\textwidth}\ \\[12pt]
 978-1-5386-3395-3/18/\$31.00 \copyright 2018 IEEE \\
\end{minipage}}
\usepackage{graphicx}
\usepackage{bbm}
\usepackage{amsfonts}
\usepackage{chngcntr}
\usepackage{times}
\usepackage{latexsym}
\usepackage{amssymb}
\usepackage{amsmath}
\usepackage{cite}
\usepackage{verbatim}
\usepackage{subfigure}
\usepackage{multirow}
\usepackage{lastpage}
\usepackage{calc}
\usepackage{eso-pic}
\usepackage{tabularx}
\usepackage{ifthen}
\usepackage{epstopdf}
\usepackage[shortlabels]{enumitem}
\usepackage{textcomp}
\usepackage{caption}
\usepackage[font={small}]{caption}

\usepackage[english]{babel}
\usepackage{textgreek}
\usepackage[utf8x]{inputenc}
\usepackage{algorithm}
\usepackage{algorithmic}

\counterwithin{cor}{theorem}

\def\bb0{{\mathbb{0}}}

\def\bb{{\mathbf{b}}}

\def\b0{{\mathbf{0}}}

\def\bbE{{\mathbb{E}}}

\def\bbP{{\mathbb{P}}}

\def\bbR{{\mathbb{R}}}

\def\cL{\mathcal{L}}

\def\sf0{{\mathsf{0}}}

\usepackage{footnote}
\allowdisplaybreaks

\ifCLASSINFOpdf
\else
\fi
\begin{document}
%
\title{Device-to-Device Communications in the Millimeter Wave Band: A Novel Distributed Mechanism}



%
\author{\IEEEauthorblockN{Niloofar Bahadori\IEEEauthorrefmark{1}, Nima Namvar\IEEEauthorrefmark{1}, Brian Kelley\IEEEauthorrefmark{2}, Abdollah Homaifar\IEEEauthorrefmark{1} }
\IEEEauthorblockA{\IEEEauthorrefmark{1} North Carolina A\&T State University }
\IEEEauthorblockA{\IEEEauthorrefmark{2} University of Texas at San Antonio
\\ Email:\{nbahador, nnamvar\}@aggies.ncat.edu}brian.kelley@utsa.edu, homaifar@ncat.edu }


\maketitle

\begin{abstract}
In spite of its potential advantages, the large-scale implementation of the device-to-device (D2D) communications has yet to be realized, mainly due to severe interference and lack of enough bandwidth in the microwave ($\mu$W) band. Recently, exploiting the millimeter wave (mmW) band for D2D communications has attracted considerable attention as a potential solution to these challenges. However, its severe sensitivity to blockage along with its directional nature make the utilization of the mmW band a challenging task as it requires line-of-sight (LOS) link detection and careful beam alignment between the D2D transceivers. In this paper, we propose a novel distributed mechanism which enables the D2D devices to discover unblocked LOS links for the mmW band communication. Moreover, as such LOS links are not always available, the proposed mechanism allows the D2D devices to switch to the $\mu$W band if necessary. In addition, the proposed mechanism detects the direction of the LOS links to perform the beam alignment. We have used tools from stochastic geometry to evaluate the performance of the proposed mechanism in terms of the signal-to-interference-plus-noise ratio (SINR) coverage probability. The performance of the proposed algorithm is then compared to the one of the single band (i.e., $\mu$W/mmW) communication. The simulation results show that the proposed mechanism considerably outperforms the single band communication.

\emph{Keywords}- Millimeter wave; D2D; Blockage; Beam alignment
\end{abstract}
\section{Introduction}
Device-to-device (D2D) communication allows user equipments to communicate over direct links rather than going through the cellular infrastructure and thereby, is envisioned to improve the spectrum efficiency by offloading the cellular network \cite{asadi2014survey}.
However, spectrum scarcity and the interference caused by the overlaid cellular transmissions have caused ubiquitous implementation of D2D communications to be halted \cite{tehrani2014device}\cite{namvar2015context}. Exploiting the higher radio frequency bands --known as the millimeter wave (mmW) band-- for D2D communications is seen as an attractive solution to address the challenges facing the large-scale D2D implementation \cite{qiao2015enabling}. The mmW band communication brings new possibilities for network planning. For instance, as unlicensed spectrum is abundant in the mmW band\cite{rappaport2013millimeter}, spectrum scarcity is no longer a serious problem. Moreover, its high path loss requires employing directional antennas, which in turn alleviates the problem of multi-user interference (MUI) \cite{niu2015survey}.

However, before reaping the potential advantages of D2D communication in the mmW band, one needs to address several new technical challenges. First, unlike the microwave ($\mu$W) band, the mmW band communication is known to be susceptible to blockage, and it also undergoes severe attenuation in non-line-of-sight (NLOS) links \cite{rappaport2013millimeter}. Consequently, an outage is more than likely to happen in NLOS/blocked links. Second, employing directional antennas requires beam alignment at both transmitter and receiver ends \cite{wei2014key}, incurring significant overhead to the system. Therefore, establishing a reliable D2D communication link at the mmW band requires devising an effective mechanism to perform a low-overhead beam alignment in order to enable directional LOS communication. Furthermore, NLOS/blocked links are required to be detected and avoided. Several approaches have been proposed in the literature to address these challenges.

In order to avoid NLOS/blocked links, authors in \cite{pi2011introduction} introduced a mechanism in which the $\mu$W band is used for transmitting the control signals, while the mmW band is utilized for data transmissions. Furthermore, to detect LOS links, a centralized reinforcement-learning based algorithm is proposed in \cite{semiari2017joint} which schedules the users on either the mmW band or the $\mu$W band based on LOS link availability. A similar model is discussed in \cite{wang2016hybrid} where the link scheduling is based on the channel information received from D2D nodes. To perform the beam alignment for directional communication in the mmW band, an exhaustive-search based algorithm is proposed in \cite{nitsche2014ieee} in order to detect the direction of the intended pair. Similar works for indoor beam alignment can be found in \cite{nitsche2015steering} and \cite{sur2016beamspy}. However, these works are mostly centralized and employ exhaustive search algorithms which impose significant overhead to the network. Moreover, their focus is mainly on either link detection or beam alignment while addressing both problems simultaneously and using a low-overhead approach is lacking in the literature.

In this paper, a novel mechanism is proposed which enables the D2D devices to select between the mmW band and the $\mu$W band for data transmission by detecting the LOS\footnote{We assume that a D2D link is LOS only if the link between D2D transmitter-receiver pair is not intersected with any blockages.} links along with their direction for proper beam alignment. Unlike the previous works, our proposed mechanism is distributed and thus, can be employed in infrastructure-less communication scenarios such as ad-hoc networks. Moreover, we employ stochastic geometry to analyze the performance of the proposed mechanism and compare it with one of the single band D2D communications. Simulation results show the proposed mechanism yields significant improvement in terms of the signal-to-interference-plus-noise ratio (SINR) coverage probability over the single band D2D communications.

The remainder of this paper is organized as follows. The system model and the proposed mechanism is described in Section \ref{sec:systemModel}. In Section \ref{sec:analysis} the performance of the proposed mechanism is analyzed, using tools from stochastic geometry. Simulation results are presented in Section \ref{sec:result} and finally, conclusions are drawn in Section \ref{sec:Conclusion}.
\setlength\belowcaptionskip{-2.45ex}
\begin{figure}
\centering
\includegraphics[ width=5.5cm, height=4.5cm, trim=1cm 5.5cm 1cm 5.6cm, clip]{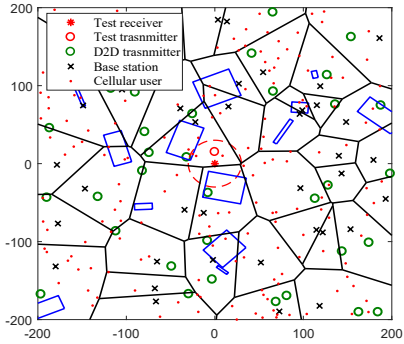}
	\caption{A sample realization of the described network model, where blue rectangles represent the building blockages.}
	\label{fig:network}
\end{figure}

\section{System Model}\label{sec:systemModel}
Consider a D2D underlaid cellular network in which D2D transmitters (DTs) are spatially distributed according to a homogeneous Poisson point process (PPP) $\mathbf{\Phi}_\text{DT}=\{d_i\}$ where $d_i \in \mathbb{R}^2$ denotes the location of $i$-th DT. Base stations (BSs) and cellular users (CUs) are also spatially distributed according to two independent PPPs $\mathbf{\mathbf{\Phi}}_\text{B}=\{b_i\}$ and $\mathbf{\Phi}_\text{C}=\{c_i\}$ with density $\lambda_\text{B}$ and $\lambda_\text{C}$, respectively. Random size rectangular building blockages are also distributed randomly by another independent PPP. Fig. \ref{fig:network} shows a sample realization of the network.
Moreover, assume that D2D users are capable of communicating in both the mmW band and the $\mu$W band. In order to avoid the high MUI in the $\mu$W band, D2D users tend to transmit their traffic in the mmW band. However, as the mmW band transmission requires a clear LOS link, the mmW band is selected only if the communication link between the D2D pair is LOS; otherwise, the $\mu$W band is used for D2D transmission. Therefore, to establish a D2D link in the mmW band, D2D devices are required to detect the LOS links (if existent) along with the direction of their corresponding pair in order to perform suitable beam alignment.

In the following sections, we elaborate on a distributed mechanism which enables the D2D devices to select between the mmW and the $\mu$W band by detecting LOS links and their direction to perform beam alignment.
\subsection{\textbf{D2D Peer's Profile}}
Suppose that all D2D devices are equipped with multiple-input-multiple-output (MIMO) antenna arrays, and constantly broadcast a peer-discovery beacon in the $\mu$W band, in order to announce their presence to their proximate peers\cite{wu2013flashlinq}. Upon receiving the peer-discovery beacon from the intended peer, D2D devices build the angle of arrival (AoA) spectrum for the received signals, by comparing the signal's phase at its multiple antennas. The AoA spectrum represents the incoming signal power as a function of the angle of incidence, i.e., $\mathcal{A}=\{I_i\measuredangle\alpha_i\}_{i=1}^{N}$ in which $I_i$ and $\alpha_i$ denote the magnitude and the angle of incidence of the $i$-th peak and $N$ denotes the total number of peaks in the AoA spectrum, $\mathcal{A}$. Note that $N$ is a random variable depending on the environmental conditions and thus, is not known a priori. Each D2D device builds a profile for its intended D2D peer by storing the AoA spectrum in subsequent time intervals. Let $\mathcal{P}\triangleq\{\mathcal{A}_j\}_{j=1}^W$ denotes the intended user's profile in which $\mathcal{A}_j$ is the AoA spectrum at the $j$-th time step and $W$ is the window size, i.e., the number of stored AoA spectrums.
\subsection{\textbf{Link Detection and Beam Alignment}}\label{sub}
In the absence of reflectors in the environment, if there exists a clear LOS link, the peak of the AoA spectrum shows the direction of the transmitter. However, if the LOS link is blocked, the peak of the AoA spectrum shows the direction of a strong reflector in the environment which caused the multipath signal to reach the receiver. The question is \emph{how to recognize if the AoA spectrum’s peak corresponds to a LOS transmitter or a random reflector?} Note that if the peak of the AoA spectrum is caused by a reflector, even small changes in the location of D2D devices (including small body movements), results in significant disparity among the AoA spectrums in the transmitter profile, while direct path peak remains relatively unchanged\cite{xiong2013arraytrack}. Consequently, in order to detect the LOS link, users build the \emph{combined AoA spectrum} by maintaining the overlapping peaks and removing the rest of the peaks. The combined AoA spectrum $\tilde{\mathcal{A}}$ can be defined as
\begingroup\makeatletter\def\f@size{9}\check@mathfonts
\begin{equation}\label{Atilda}
  \tilde{\mathcal{A}}=\bigg\{I_k \measuredangle\alpha_k \bigg| \sum_{j=1}^{W}\mathbbm{1}(\alpha_k \in \mathcal{A}_j)= W \bigg\},
\end{equation}
in which $\mathbbm{1}(.)$ is the indicator function which is equal to $1$ if its argument holds true and is $0$ otherwise. Moreover, the magnitude $I_k$ that is the average of all peaks with angle of incidence $\alpha_k$ is given by
\begin{equation}\label{Amplitude}
  I_k=\frac{1}{W}\sum_{j=1}^{W}\sum_{n=1}^{|\mathcal{A}_j|} \mathbbm{1}(\alpha_n=\alpha_k)I_n,
\end{equation}
where $|\mathcal{A}_j|$ represents the cardinality of the set $\mathcal{A}_j$.

Having built the combined AoA spectrum, D2D devices detect LOS links along with their direction and select the suitable frequency band for data transmission. If $|\tilde{\mathcal{A}}|=1$, i.e., the combined AoA spectrum has a single peak, the LOS link exists and thus, the mmW band is selected for D2D data transmission. Moreover, its angle of incidence shows the direction the intended peer that will be used for beam alignment. However, if $|\tilde{\mathcal{A}}| \neq 1$,
then there are either multiple peaks or no peak in the AoA spectrum which corresponds to having NLOS link. Hence the $\mu$W band is selected. Fig. \ref{fig:Aoa} shows an example of user's profile with $W=2$ along with its combined AoA spectrum.

Next, we develop a stochastic geometric framework to analyze the performance of the system model. 

\begin{figure}
\centering
\includegraphics[width=.80\columnwidth,  trim=2cm 8.6cm 1cm 9.6cm, clip]{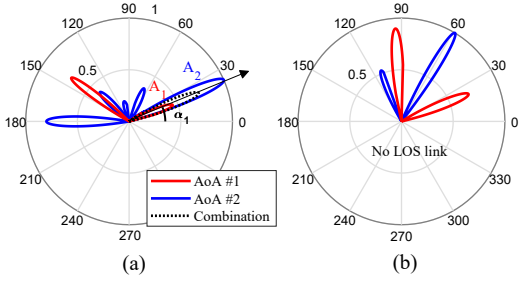}
	\caption{AoA profile with $W=2$: (a) LOS link is available and $\alpha_1$ is the direction of intended peer, (b) no LOS link is available.}
\setlength{\abovecaptionskip}{-3cm}
	\label{fig:Aoa}
\end{figure}

\section{Analysis}\label{sec:analysis}
In order to analyze the performance of the proposed mechanism, we assume that the mmW band spectrum is dedicated for D2D communication. In the $\mu$W band, D2D users are underlaid the cellular network and perform spectrum sensing to opportunistically access the $\mu$W downlink resources. As mentioned before, locations of the network elements are modeled by independent homogeneous PPPs, due to their tractability. It is shown in \cite{bai2014analysis} that using a random boolean scheme of rectangles to model the blockages, a link of length $\|x\|$ is LOS with probability $p_\text{\tiny LOS}(\|x\|)=\exp(-\beta \|x\|)$, where parameter $\beta$ depends on the average size and density of blockages. Thus, the probability of NLOS link is defined as $p_\text{\tiny NLOS}(\|x\|) = 1-p_\text{\tiny LOS}(\|x\|)$.

Assume that all DTs and BSs are transmitting at a constant transmit power, $P_D$ and $P_B$, respectively. Each communication link experiences i.i.d small-scale Rayleigh fading. Hence, the received signal power can be modeled as an exponential random variable with parameter 1.

Here, we use the SINR coverage probability as a metric to assess the performance of the network.
The SINR coverage probability is defined as the probability that the received SINR is higher than a predefined threshold $\gamma$, i.e., $p_{\text{c}} (\gamma)=\bbP[\text{SINR}\geq\gamma]$.
The performance metric is obtained for a \emph{test D2D receiver} at the origin $(0,0)\in \mathbb{R}^2$, while the results hold for any generic receiver, based on the Slivnyak's theorem \cite{baccelli2010stochastic}.

For the test D2D receiver, the received SINR is defined as
\begin{equation}\label{eq:SINR}
  \text{SINR}_i=\frac{P_D h_0 G_e \text{PL}(d_0)}{\sigma^2+I_i},
\end{equation}
where $i\in \{mm,\mu\}$ represents the transmission band (mmW/$\mu$W) and $h_0$ is the channel gain. $G_e= G_{\text{Tx}_0}G_{\text{Rx}_0}$ denotes the effective antenna gain at the test receiver, in which $G_{\text{Tx}_0}$ and $G_{\text{Rx}_0}$ are transmitter's and receiver's antenna gain, respectively. $\text{PL}(d_0)= C d_0^{-\alpha}$ denotes the distance dependent path loss model, in which $d_0$ is the link's length and $C={(\frac{\lambda}{4\pi})}^2$ where $\lambda$ is the wavelength, and $\alpha$ is the path loss exponent. Finally, $\sigma^2$ represents the noise power and $I_i$ denotes the aggregate interference.

Remember that using the proposed mechanism in Section \ref{sec:systemModel}, D2D devices transmit over the mmW band when there exists a LOS link, otherwise the $\mu$W band is exploited. Hence, the SINR coverage probability for the test receiver is given by
\begin{align}\allowdisplaybreaks
p_\text{c}(\gamma)
&=p_\text{c}^{\text{mm}}(\gamma) p_{\text{LOS}}(d_0) + p_\text{c}^{\mu}(\gamma) p_{\text{NLOS}}(d_0)\label{eq:hybrid}, 
\end{align}
where $p_\text{c}^{\text{mm}}(\gamma)$ and $p_\text{c}^{\mu}(\gamma)$ denote the test receiver's SINR coverage probability in the mmW band and the $\mu$W band, respectively.

Next, the performance of D2D network in the mmW band and the $\mu$W band is analyzed to obtain the network performance.
 \begin{figure}
\centering
\includegraphics[width=.48\columnwidth, trim=0cm 0cm 0cm .20cm, clip]{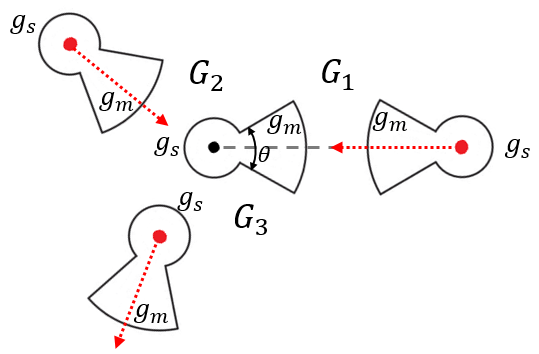}
	\caption{the mmW band D2D transceivers' antenna: the black node shows the test receiver and the red nodes depict the DTs.}
	\label{fig:directional interference}
\end{figure}
\subsection{\textbf{D2D Communication in the mmW Band}}\label{sub:mmW}
In the mmW band, no prior coordination among devices for interference mitigation is assumed. Suppose that, each DT has an intended receiver in its coverage area, and also at least one packet ready for transmission. The DTs access the entire mmW spectrum based on the slotted Aloha protocol, with access probability $q_a$. The simplified sectored model, as shown in Fig. \ref{fig:directional interference}, is used to model the steerable and directional mmW antenna array \cite{bai2015coverage}. The antenna pattern of D2D devices in the mmW band are modeled by two parameters, namely, $g_\text{m}$
for the mainlobe gain with beamwidth $\theta$, and $g_\text{s}$
for sidelobe gain with beamwidth $2\pi-\theta$.

As we have no prior information about the antenna direction $\varphi$ at different D2D transmitters, we assume that $\varphi$ is modeled as a uniform random variable $\varphi\sim \mathcal{U}(0,2\pi)$.
Given the simplified sectored model explained above, the effective antenna gain from a DT located at $d_i$ to the test receiver at the origin can be defined as a discrete random variable \cite{bai2015coverage}
\begin{equation}
G_e =
\begin{cases}
G_1 = g_{\text{m}}g_{\text{m}} & \small\text{with}  \hspace{2mm} p_1=p^2 \\
G_2 = g_{\text{m}}g_{\text{s}} & \small\text{with}  \hspace{2mm} p_2=2p(1-p) \\
G_3 = g_{\text{s}}g_{\text{s}} & \small\text{with}  \hspace{2mm} p_3=(1-p)^2, \\
\end{cases}\label{eq:ant_gain2}
\end{equation}
where $p=\frac{\theta}{2\pi}$ denotes the antenna mainlobe's angular coverage probability.

Since in the mmW band only LOS links are utilized for D2D communication, the test receiver's SINR is defined as 
\begin{align}
\text{SINR}_\text{mm} &= \frac{P_D h_0 G_e C d_0^{-\alpha_\text{L}}}{\sigma^2+I_\text{mm}},\label{eq:SINR_mmW}
\end{align}
where $\alpha_\text{L}$ represents the LOS path loss exponent. $G_e=g_{\text{m}}g_{\text{m}}$, since perfect beam alignment is considered between corresponding D2D pairs. $I_\text{mm}$ denotes the aggregate interference in the mmW band and is defined as
\begin{align}\allowdisplaybreaks
I_\text{mm} = \sum_{\text{j}=1}^{3}\sum_{d_i \in \tilde{\mathbf{\Phi}}_{\text{DT}_j}} P_D h_{d_i} G_j  \text{PL}(\|d_i\|),\label{eq:interference_mm}
\end{align}
where $\|.\|$ denotes the Euclidian distance.
Based on the channel access probability $q_a$, and the effective antenna gain in \eqref{eq:ant_gain2}, PPP distribution of DTs, $\mathbf{\Phi}_{\text{DT}}$, can be thinned into three independent PPPs $\tilde{\mathbf{\Phi}}_{\text{DT}_j}$,  $j \in \{1,2,3\}$ with intensity
$\lambda_{\tilde{\mathbf{\Phi}}_{\text{DT}_j}}=q_\text{a} p_\text{j} \lambda_{\text{DT}}$. In this work, NLOS interference is neglected due to its negligible effect on the interference distribution.

Using \eqref{eq:SINR_mmW} and \eqref{eq:interference_mm}, the test receiver's SINR coverage probability in the mmW is defined as
\begin{align}\allowdisplaybreaks
p_\text{c}^{\text{mm}}(\gamma) &=\bbP\left[\text{SINR}_\text{mm}\geq\gamma\right]\nonumber\\
&= \bbP\bigg[h_0 \geq \varepsilon_{\text{L}}(\sigma^2 + I_\text{mm}) \bigg] p_{\text{LOS}}\left(d_0\right) \nonumber \\
&= \bbE_{h,\tilde{\mathbf{\Phi}}_{\text{DT}_j}}\bigg [\exp\left(-\varepsilon_{\text{L}}(\sigma^2 + I_\text{mm})\right)\bigg]\exp(-\beta d_0)\label{eq:laplace_mm}\\
&=\exp(-\varepsilon_{\text{L}}\sigma^2)\cL_{I_{\text{mm}}}(\varepsilon_{\text{L}}) \exp(-\beta d_0)\label{eq:final_mm},
\end{align}
where $\varepsilon_{\text{L}}=\frac{\gamma d_0^{\alpha_{\text{L}}}}{P_Dg_\text{m}g_\text{m} C}$. Equation \eqref{eq:laplace_mm} follows due to the exponential distribution of the channel gain, $h_0$.
Notice that $\bbE_{h,\tilde{\mathbf{\Phi}}_{\text{DT}_j}}\left[\exp(-\varepsilon_\text{L} I_{\text{mm}})\right]$
corresponds to the Laplace transform of the aggregate interference $I_\text{mm}$ and can be written as
\begin{align}\allowdisplaybreaks
\cL_{I_{\text{mm}}}(\varepsilon_{\text{L}}) &= \prod_{\text{j}=1}^{3}\exp\left\{\int_{0}^
{\infty}\big((1+ C_{\text{L}}G_\text{j} r^{-\alpha_{\text{L}}})^{-1}-1\big) \lambda_{\tilde{\mathbf{\Phi}}_{\text{DT}_j}}(r)\text{d}r\right\}\label{eq:laplace_Imm},
\end{align}
where $ \lambda_{\scriptstyle\tilde{\mathbf{\Phi}}_{\text{DT}_j}}(r)=2\pi r q_\text{a} p_\text{j} \exp(-\beta r)\lambda_{\mathbf{\Phi}_{\text{DT}}}$, and $C_{\text{L}}=P_D C \varepsilon_{\text{L}}$.\\

\begin{IEEEproof}
See the Appendix \ref{app:proof}.
\end{IEEEproof}
\vspace{3mm}
Next, we explain how D2D users underlaid the downlink cellular network access the cellular BSs' resources for data transmission.
\subsection{\textbf{D2D Communication in the $\mu$W Band}}\label{sub:microwave}
Suppose that in the $\mu$W band, the available downlink spectrum is divided into $K$ orthogonal frequency channels, out of which one channel, denoted by $k_d$, can be used for D2D transmissions. Each CU is assigned to its nearest BS, and is served only by one channel. Note that $k_d$ is not exclusive for D2D communication; however, BS utilizes this channel only if all other channels are occupied. The probability that $k_d$ is used by a generic BS, denoted by $p_{k_d}$, depends on the distribution of number of CUs associated with that BS. $p_{k_d}$ is calculated in \cite{elsawy2013cognitive} for the network in which BSs and CUs are spatially distributed based on independent PPPs.

In order to mitigate the interference caused by the nearby BS transmissions on the D2D channel $k_d$, the cognitive D2D model in \cite{sakr2015cognitive} is used, in which DTs sense the state of the channel $k_d$ before using it. In this case, DTs use the $k_d$ only if the amount of the received interference from all BSs that are using the D2D channel $k_d$ is less than a sensing threshold, $\tau$. The cognition forms a circular threshold region around each D2D user that guarantee no BS uses channel $k_d$ in this region. The radius of the threshold region, can be defined as $d_{\tau} = (\frac{P_B  h_b}{\tau})^{\delta}$,
where $\delta=\frac{1}{\alpha_{\mu}}$, $\alpha_{\mu}$ is the path loss exponent and $h_b$ denotes the channel gain from the BS located at $b$.

The average radius of the threshold region is expressed as
\begin{equation}\label{eq: Threshold region}
\bar{d_{\tau}} = \left(\frac{ P_B }{\tau}\right)^{\delta}\bbE_h[h_b^{\delta}],
\end{equation}
where $\bbE_h[h_b^{\delta}]=\int_{0}^{\infty}h_b^{\delta}\exp(-h_b)\text{d}h_b =\Gamma(1+\delta)$ due to the exponential distribution of $h_b$ and $\Gamma(.)$ is the gamma function.

Using the thinning property of a PPP \cite{baccelli2010stochastic}, the spatial distribution of BSs that use D2D channel $k_d$ for transmission, forms a PPP $\tilde{\mathbf{\Phi}}_{\text{B}}$ with density $p_{kd}\lambda_{\text{B}}$. Thus, number of BSs that use $k_d$ in the threshold region of a generic D2D user has Poisson distribution. Therefore, the probability that channel $k_d$ is not used by any BS in the threshold region i.e., the $k_d$ is available for D2D transmission is defined as 
\begin{equation}\label{eq:available probability}
  p_a=\exp\left(-\lambda_\text{B} p_{k_d} \pi\bar{d_{\tau}}^2\right),
\end{equation}
where $\pi\bar{d_{\tau}}^2$ denotes the average area of the threshold region.

The test receiver's SINR in the $\mu$W band is defined as
\begin{equation}\label{eq:SINR_micowavwe}
\text{SINR}_\mu=\frac{P_D h_{\text{0}} G_e C d_0^{-\alpha_{\mu}}}{\sigma^2+I_{\mu}},
\end{equation}
where $\alpha_\mu$ is the path loss exponent. We assume $G_e =1$, since no beamforming is considered in the $\mu$W band. $I_\mu$ denotes the aggregate interference in the $\mu$W band and is defined as
\begin{align}\allowdisplaybreaks
 I_{\mu} &= \overbrace{\sum_{d_i \in \tilde{\mathbf{\Phi}}_{\text{DT}}} P_D h_{d_i} C\|d_i\|^{-\alpha_{\mu}}}^{I_{\text{DT}}}+\overbrace{\sum_{b_i \in \tilde{\mathbf{\Phi}}_{\text{B}}} P_B h_{b_i} C \|b_i\|^{-\alpha_{\mu}}}^{I_{\text{BS}}},\label{eq: agg interference}
\end{align}
where $ \tilde{\mathbf{\Phi}}_{\text{DT}}$ denotes the spatial distribution of DTs that use channel $k_d$ for transmission, which is a PPP with intensity $\lambda_{\tilde{\mathbf{\Phi}}_{\text{DT}}}= p_a \lambda_{\text{DT}}$, and $\tilde{\mathbf{\Phi}}_{\text{B}}$ denotes the distribution of BSs that are using $k_d$ outside the threshold region.
\begin{table}
\caption{Simulation Parameters}
    \centering
       \begin{tabular}{|c|c|c|c}
       \hline
       Parameter&Notation& Value \\ \hline\hline\hline
       BS/DT power & $P_B$, $P_{\text{D}}$ & $37$, $0$ (dBm) \\
       Antenna gain & $g_m$, $g_s$ & $10$, $-10$ (dBi)  \\
       Mainlobe beamwidth & $\theta$&  $30^{\circ}$\\
       Density of PPPs & $\lambda_{\text{B}}$, $\lambda_{\text{C}}$, $\lambda_{\text{DT}}$ & $1, 5, 50$ ($\text{km}^{-2}$)\\
       Path-loss exponent  & $\alpha_{\mu}$, $\alpha_{\text{L}}$, $\alpha_{\text{N}}$& $4$, $2$, $5$ \\
       Interference threshold& $\tau$ & $-85$ (dBm)\\
       Bandwidth & $B_{\mu}$, $B_{\text{mm}}$ & $0.1$, $1$ (GHz)\\
       Carrier frequency& $f_{\mu}$, $f_{\text{mm}}$ & $2$, $28$ (GHz)\\
       Noise power & $\sigma^2$ & $-174+10 \log_{10}B_i+10$ {\footnotesize(dBm)} \\
       \hline
       \end{tabular}\label{params}
\end{table}
Using \eqref{eq: agg interference} and \eqref{eq:SINR_micowavwe}, the test receiver's SINR coverage probability is defined as
\begin{align}\allowdisplaybreaks
p_\text{c}^{\mu}(\gamma) &=\bbP\left[\text{SINR}_\mu\geq\gamma\right]\nonumber\\
&= \bbP\bigg[h_0 \geq \varepsilon \big(\sigma^2 +I_{\text{DT}} + I_{\text{BS}}\big)\bigg]\nonumber\\
&= \bbE_{h,\tilde{\mathbf{\Phi}}_\text{B},\tilde{\mathbf{\Phi}}_{\text{DT}}}\bigg [\exp\big(-\varepsilon(\sigma^2+ I_{\text{DT}}+ I_{\text{BS}})\big)\bigg]\label{eq:Laplace} \\
&= \exp(-\varepsilon \sigma^2) \cL_{I_{\text{DT}}}(\varepsilon)\cL_{I_{\text{BS}}}(\varepsilon)\label{eq:laplace_uw},
\end{align}
where $\varepsilon=\frac{\gamma d_0^{\alpha_{\mu}}}{CP_D}$. Equation \eqref{eq:Laplace} follows due to the exponential distribution of the channel gain. Notice that $\bbE_{h,\tilde{\mathbf{\Phi}}_{\text{DT}}}[\exp(-\varepsilon I_{\text{DT}})]$ and $\bbE_{h,\tilde{\mathbf{\Phi}}_\text{B}}[\exp(-\varepsilon I_{\text{BS}})]$ corresponds to the Laplace transform of $I_\text{BS}$ and $I_{\text{DT}}$, respectively. Laplace transforms can be written as
\begin{align}\allowdisplaybreaks
 \cL_{I_{\text{DT}}}(\varepsilon)&=  \exp \left\{-2 p_a \lambda_{\text{DT}} \varepsilon_{\text{DT}} ^{\delta}\frac{\pi^2 \delta}{\sin(2 \pi \delta)}\right\}\label{eq:laplac_mic_DT}\\
  \cL_{I_\text{BS}}(\varepsilon)&= \exp \bigg\{-\pi p_{k_d} \lambda_\text{B} \sqrt{\varepsilon_{\text{B}} } \big(\frac{\pi}{2}-\text{tan}^{-1}(\frac{1}{\vartheta})\nonumber \\
&\mathrel{\phantom{=}} +\frac{\vartheta}{\vartheta^2+1} \big)+\frac{p_{k_d}\lambda_\text{B} \pi \varepsilon_{\text{B}} \bar{d_{\tau}}^2}{\varepsilon_{\text{B}} + \bar{d_{\tau}}^{\alpha_{\mu}}} \bigg\}\label{eq:laplaceBS},
\end{align}
where $\varepsilon_{\text{DT}}= C P_D\varepsilon$, $\varepsilon_{\text{B}}=CP_B\varepsilon $, $\vartheta = \sqrt{\varepsilon_{\text{B}} \bar{d_{\tau}}^{-\alpha_{\mu}}}$ and
$\delta=\frac{1}{4}$. Equation \eqref{eq:laplac_mic_DT} is derived the same as \eqref{eq:laplace_Imm}, and \eqref{eq:laplaceBS} is derived using (3.46) in \cite{baccelli2010stochastic}.

 \begin{figure}
\centering
\includegraphics[width=.78\columnwidth, trim=.7cm 6.1cm 1.5cm 7.1cm, clip]{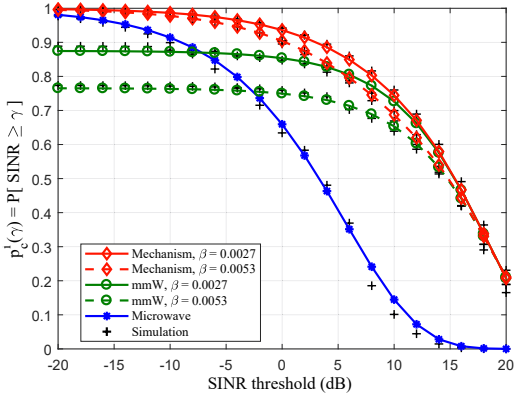}
\caption{D2D network's SINR coverage probability vs. SINR threshold with $d_0$ $= 50$ $m$, $\lambda_{\text{DT}}=50$ $km^{-2}$.}
	\label{fig:Coverage_SINR}
\end{figure}
\section{Numerical Results and Discussions}\label{sec:result}
Using the coverage probability formulas in \eqref{eq:hybrid}, \eqref{eq:final_mm} and \eqref{eq:laplace_uw} as the performance metrics, the performance of the proposed mechanism is compared to single band (i.e., mmW/$\mu$W) D2D communications. Moreover, to validate our analytical results, we simulated a network similar to the one discussed in the system model. For our simulations, we consider an area of the size $10$ $km$ $\times$ $10$ $km$  which is --given the transmit power of D2D devices-- large enough to avoid the boundary effect. D2D transmitters along with various size rectangular blockages are distributed in the area according to PPP. Also, we assume that all the transmitters use a constant power for transmission. Table \ref{params} summarizes the simulation parameters. To thwart the effect of noisy data, we used Monte Carlo simulation with $10,000$ iterations and averaged out the results. In the following figures, simulation results are represented by "$+$" symbol.

Fig. \ref{fig:Coverage_SINR} shows the SINR coverage probability of D2D network as a function of the SINR threshold, with two different blockage densities, namely, $\beta = 0.0027$ and $\beta=0.0053$.
It shows that by increasing the density of blockages, the SINR coverage probability of D2D receivers in the mmW band decreases. It is in agreement with the observation that increasing the number of blockages in the environment, lowers the chance of LOS links, and thereby, decreases the SINR coverage probability. Moreover, it is seen that the proposed mechanism improves network performance by about $30\%$ at $\gamma = 0$ $dB$, compared to D2D communication in the $\mu$W band. Finally, it shows that the simulation results closely follow the analytical results.


 \begin{figure}
\centering
\includegraphics[width=.83\columnwidth, trim=.7cm 6.1cm 1.5cm 7.4cm, clip]{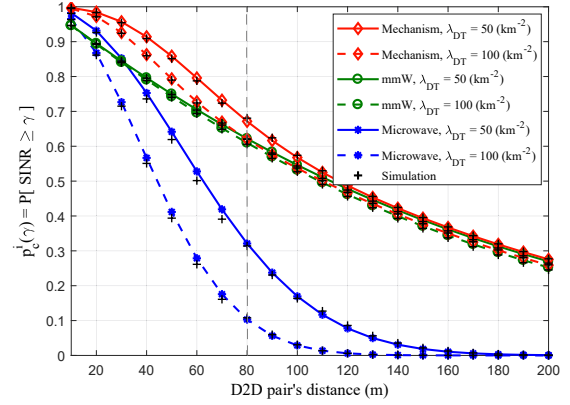}
\caption{D2D network's received SINR coverage probability vs. D2D pair distance with $\gamma = 0$ dB $\beta=0.0053$.}
	\label{fig:Coverage_density}
\end{figure}

Fig. \ref{fig:Coverage_density} shows the SINR coverage probability of the D2D network as a function of corresponding D2D pair's distance, for two different densities of the D2D transmitters, namely, $\lambda_{\text{DT}} = 50$ $km^{-2}$ and $\lambda_{\text{DT}} = 100$ $km^{-2}$. It is shown that, increasing the distance of D2D pairs, degrades the performance of the D2D network in both the mmW and the $\mu$W bands. In the $\mu$W band, increasing the distance drops the network performance even more, due to the lack of beamforming and directional communication. This figure also highlights the low MUI in the mmW band. As it can be seen, due to the directional nature of communication in the mmW band, increasing the density of interferers does not affect D2D communication's performance significantly. The proposed system model manages to improve the network performance, in particular when the distance of D2D pairs is less than $80$ $m$.

Fig. \ref{fig:rate} shows the rate coverage probability of the D2D network defined as $p_R(T)=\bbP[\text{Rate}\geq T]$, in which the rate is given by $B\log(1+\text{SINR})$, where $B$ is the system bandwidth. It is seen that rate coverage probability for the mmW band is almost constant and independent of rate thanks to its large bandwidth, while for the $\mu$W band, it is a decreasing convex function of rate, although more consistent. Our proposed mechanism inherits the merits of the communication in both spectrum bands by allowing the D2D devices to switch between them. It shows that at low rate regimes, our proposed algorithm tends to exploit the $\mu$W band as it provides a higher coverage probability for the users, while it switches to the mmW band when the rate demand increases. Overall, Fig. \ref{fig:rate} shows that the proposed mechanism outperforms the single band communication in terms of rate coverage probability.
\begin{figure}
\centering
\includegraphics[width=.80\columnwidth, trim=.7cm 7cm 1.5cm 6.8cm, clip]{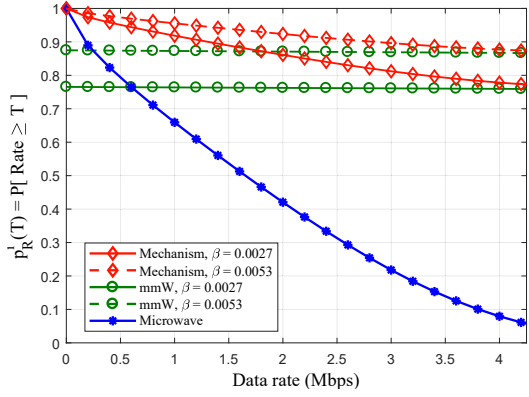}
\caption{Rate coverage probability vs. achievable rate with $d_0$ $= 50$ $m$ and $\lambda_{\text{DT}}=50$ $km^{-2}$.}
	\label{fig:rate}
\end{figure}
\section{Conclusion}\label{sec:Conclusion}
In this paper, we proposed a novel distributed mechanism which enables D2D devices to select between the mmW band and the $\mu$W band for data transmission. In order to devise a distributed mmW band communication protocol for D2D communications, the D2D users are in charge of detecting LOS links along with their corresponding direction to perform proper beam alignment. Our proposed algorithm enables the D2D devices to perform such a task by using peer-discovery beacons and comparing the AoA spectrum of their intended peer over subsequent time intervals. We have used stochastic geometry to provide a complete framework to analyze the performance of the proposed mechanism in terms of the received SINR coverage probability of D2D users for which closed-form analytical formulas are derived. Our simulation results demonstrate that the proposed mechanism achieves considerable performance gain over the single band (i.e., mmW/$\mu$W) D2D communications. Moreover, our simulations validate the analytical results discussed in the paper.

\appendix\label{app:proof}
2D PPP $\tilde{\mathbf{\Phi}}_{\text{DT}_j}$ is mapped onto $\bbR^+$ by letting $\mathbf{\Phi}_\text{j}=\{\|d_i\|=r_i\}$, be the distances of points of the PPP $\tilde{\mathbf{\Phi}}_{\text{DT}_j}$ that are LOS to the test receiver, with density  $\lambda_{\mathbf{\Phi}_\text{j}}(r)= 2 \pi r p_\text{LOS}(r)\lambda_{\tilde{\mathbf{\Phi}}_{\text{DT}_j}}$. Using \eqref{eq:interference_mm}, the Laplace transform of $I_\text{mm}$ can be defined as
\begingroup\makeatletter\def\f@size{8}\check@mathfonts
\begin{align}\allowdisplaybreaks\label{eq:proof}
\cL_{I_{\text{mm}}}(\varepsilon_{\text{L}})
&=\bbE_{h,\tilde{\mathbf{\Phi}}_\text{j}}\left[\exp(-\varepsilon_{\text{L}}I_{\text{mm}})\right]\nonumber\\
&= \bbE_{h,\tilde{\mathbf{\Phi}}_\text{j}}\left[\exp(-\varepsilon_\text{L} \sum_{j=1}^{3} \sum_{r \in \tilde{\mathbf{\Phi}}_\text{j}}  P_D C  G_\text{j} h_r r^{-\alpha_\text{L}}\right]\nonumber\\
&= \bbE_{h,\tilde{\mathbf{\Phi}}_\text{j}}\left[\prod_{\text{j}=1}^{3}\prod_{r \in \tilde{\mathbf{\Phi}}_\text{j}}\exp(- C_\text{L} G_\text{j} h_{r} r^{-\alpha_\text{L}})\right]\\
&= \bbE_{\tilde{\mathbf{\Phi}}_\text{j}}\left[\prod_{\text{j}=1}^{3}\prod_{r \in \tilde{\mathbf{\Phi}}_\text{j}}\bbE_{h}\left[\exp(- C_\text{L} G_\text{j} h_{r} r^{-\alpha_\text{L}})\right]\right]\label{eq:iid}\\
&= \prod_{\text{j}=1}^{3}\exp \left\{-\int_{0}^{\infty}\bbE_{h}\left[1-\exp(- C_\text{L} G_\text{j} h_{r}  r^{-\alpha_\text{L}})\right]\lambda_{\tilde{\mathbf{\Phi}}_\text{j}}(r)\text{d}r\right\}\label{eq:mgf_poisson}\\
&= \prod_{\text{j}=1}^{3}\exp \bigg\{\int_{0}^{\infty}\left((1+ C_\text{L} G_\text{j} r^{-\alpha_\text{L}})^{-1}-1\right)\lambda_{\tilde{\mathbf{\Phi}}_\text{j}}(r)\text{d}r\bigg\},\label{eq:mgf_exp}
\end{align}\endgroup
where equation \eqref{eq:iid} follows since channel gains are i.i.d, and also PPP $\tilde{\mathbf{\Phi}}_\text{j}$ and $h_r$ are independent.
Equations \eqref{eq:mgf_poisson} and \eqref{eq:mgf_exp} are derived using the probability generating functional (PGFL) of PPP $\tilde{\mathbf{\Phi}}_\text{j}$ with density $\lambda_{\mathbf{\Phi}_\text{j}}(r)$, and PGFL of $h_r$ with exponential distribution, respectively.


\section*{Acknowledgment}
The authors would like to acknowledge the support from
Air Force Research Laboratory and OSD for sponsoring this
research under agreement number FA8750-15-2-0116.



%

\medskip

\bibliographystyle{ieeetr}
\bibliography{bib}

\begin{thebibliography}{10}

\bibitem{asadi2014survey}
A.~Asadi, Q.~Wang, and V.~Mancuso, ``A survey on device-to-device communication
  in cellular networks,'' {\em IEEE Communications Surveys \& Tutorials},
  vol.~16, no.~4, pp.~1801--1819, 2014.

\bibitem{tehrani2014device}
M.~N. Tehrani, M.~Uysal, and H.~Yanikomeroglu, ``Device-to-device communication
  in 5g cellular networks: challenges, solutions, and future directions,'' {\em
  IEEE Communications Magazine}, vol.~52, no.~5, pp.~86--92, 2014.

\bibitem{namvar2015context}
N.~Namvar, N.~Bahadori, and F.~Afghah, ``Context-aware d2d peer selection for
  load distribution in lte networks,'' in {\em Signals, Systems and Computers,
  2015 49th Asilomar Conference on}, pp.~464--468, IEEE, 2015.

\bibitem{qiao2015enabling}
J.~Qiao, X.~S. Shen, J.~W. Mark, Q.~Shen, Y.~He, and L.~Lei, ``Enabling
  device-to-device communications in millimeter-wave 5g cellular networks,''
  {\em IEEE Communications Magazine}, vol.~53, no.~1, pp.~209--215, 2015.

\bibitem{rappaport2013millimeter}
T.~S. Rappaport, S.~Sun, R.~Mayzus, H.~Zhao, Y.~Azar, K.~Wang, G.~N. Wong,
  J.~K. Schulz, M.~Samimi, and F.~Gutierrez, ``Millimeter wave mobile
  communications for 5g cellular: It will work!,'' {\em IEEE access}, vol.~1,
  pp.~335--349, 2013.

\bibitem{niu2015survey}
Y.~Niu, Y.~Li, D.~Jin, L.~Su, and A.~V. Vasilakos, ``A survey of millimeter
  wave communications (mmwave) for 5g: opportunities and challenges,'' {\em
  Wireless Networks}, vol.~21, no.~8, pp.~2657--2676, 2015.

\bibitem{wei2014key}
L.~Wei, R.~Q. Hu, Y.~Qian, and G.~Wu, ``Key elements to enable millimeter wave
  communications for 5g wireless systems,'' {\em IEEE Wireless Communications},
  vol.~21, no.~6, pp.~136--143, 2014.

\bibitem{pi2011introduction}
Z.~Pi and F.~Khan, ``An introduction to millimeter-wave mobile broadband
  systems,'' {\em IEEE communications magazine}, vol.~49, no.~6, 2011.

\bibitem{semiari2017joint}
O.~Semiari, W.~Saad, and M.~Bennis, ``Joint millimeter wave and microwave
  resources allocation in cellular networks with dual-mode base stations,''
  {\em IEEE Transactions on Wireless Communications}, 2017.

\bibitem{wang2016hybrid}
F.~Wang, H.~Wang, H.~Feng, and X.~Xu, ``A hybrid communication model of
  millimeter wave and microwave in d2d network,'' in {\em Vehicular Technology
  Conference, 2016 IEEE 83rd}, pp.~1--5, IEEE, 2016.

\bibitem{nitsche2014ieee}
T.~Nitsche, C.~Cordeiro, A.~B. Flores, E.~W. Knightly, E.~Perahia, and J.~C.
  Widmer, ``Ieee 802.11 ad: directional 60 ghz communication for
  multi-gigabit-per-second wi-fi,'' {\em IEEE Communications Magazine},
  vol.~52, no.~12, pp.~132--141, 2014.

\bibitem{nitsche2015steering}
T.~Nitsche, A.~B. Flores, E.~W. Knightly, and J.~Widmer, ``Steering with eyes
  closed: mm-wave beam steering without in-band measurement,'' in {\em Computer
  Communications (INFOCOM), 2015 IEEE Conference on}, pp.~2416--2424, IEEE,
  2015.

\bibitem{sur2016beamspy}
S.~Sur, X.~Zhang, P.~Ramanathan, and R.~Chandra, ``Beamspy: Enabling robust 60
  ghz links under blockage.,'' in {\em NSDI}, pp.~193--206, 2016.

\bibitem{wu2013flashlinq}
X.~Wu, S.~Tavildar, S.~Shakkottai, T.~Richardson, J.~Li, R.~Laroia, and
  A.~Jovicic, ``Flashlinq: A synchronous distributed scheduler for peer-to-peer
  ad hoc networks,'' {\em IEEE/ACM Transactions on Networking (TON)}, vol.~21,
  no.~4, pp.~1215--1228, 2013.

\bibitem{xiong2013arraytrack}
J.~Xiong and K.~Jamieson, ``Arraytrack: A fine-grained indoor location
  system,'' Usenix, 2013.

\bibitem{bai2014analysis}
T.~Bai, R.~Vaze, and R.~W. Heath, ``Analysis of blockage effects on urban
  cellular networks,'' {\em IEEE Transactions on Wireless Communications},
  vol.~13, no.~9, pp.~5070--5083, 2014.

\bibitem{baccelli2010stochastic}
F.~Baccelli, B.~B{\l}aszczyszyn, {\em et~al.}, ``Stochastic geometry and
  wireless networks: Volume ii applications,'' {\em Foundations and
  Trends{\textregistered} in Networking}, vol.~4, no.~1--2, pp.~1--312, 2010.

\bibitem{bai2015coverage}
T.~Bai and R.~W. Heath, ``Coverage and rate analysis for millimeter-wave
  cellular networks,'' {\em IEEE Transactions on Wireless Communications},
  vol.~14, no.~2, pp.~1100--1114, 2015.

\bibitem{elsawy2013cognitive}
H.~ElSawy and E.~Hossain, ``On cognitive small cells in two-tier heterogeneous
  networks,'' in {\em Modeling \& Optimization in Mobile, Ad Hoc \& Wireless
  Networks (WiOpt), 2013 11th International Symposium on}, pp.~75--82, IEEE,
  2013.

\bibitem{sakr2015cognitive}
A.~H. Sakr and E.~Hossain, ``Cognitive and energy harvesting-based d2d
  communication in cellular networks: Stochastic geometry modeling and
  analysis,'' {\em IEEE Transactions on Communications}, vol.~63, no.~5,
  pp.~1867--1880, 2015.

\end{thebibliography}

\end{document}